\documentclass{moriond}

\usepackage{amsmath}
\usepackage{amsfonts}
\usepackage{amssymb}
\usepackage{calc}
\usepackage{mathtools}
\usepackage{subfig}
\usepackage{mathabx}
\usepackage{cancel}
\usepackage[dvipsnames]{xcolor}
\usepackage{comment}
\usepackage{graphicx}
\graphicspath{{images/}}
\usepackage{pgfplotstable}
\usepackage{soul}
\usepackage{slashed} 

\newcommand{\bell}{\boldsymbol{\ell}}

\newcommand{\be}{\begin{equation}}
\newcommand{\ee}{\end{equation}}
\newcommand{\beq}{\begin{equation}}
\newcommand{\eeq}{\end{equation}}
\newcommand{\bea}{\begin{eqnarray}}
\newcommand{\eea}{\end{eqnarray}}

\newcommand{\bL}{{\mathbf  L}}
\newcommand{\bn}{{\mathbf  n}}

\newcommand{\bV}{{\mathbf V}}

\renewcommand{\bell}{\boldsymbol{\ell}}

\newcommand{\bnabla}{{\boldsymbol{\nabla}}}

\newcommand{\dd}{\partial}

\newcommand{\HH}{{\cal H}}

\newcommand{\lan}{{\langle}}
\newcommand{\ran}{{\rangle}}

\newcommand{\De}{\Delta}

\def\be{\begin{equation}}
\def\ee{\end{equation}}
\def\bea{\begin{eqnarray}}
\def\eea{\end{eqnarray}}

\newcommand{\Photo}{\includegraphics[height=30mm]{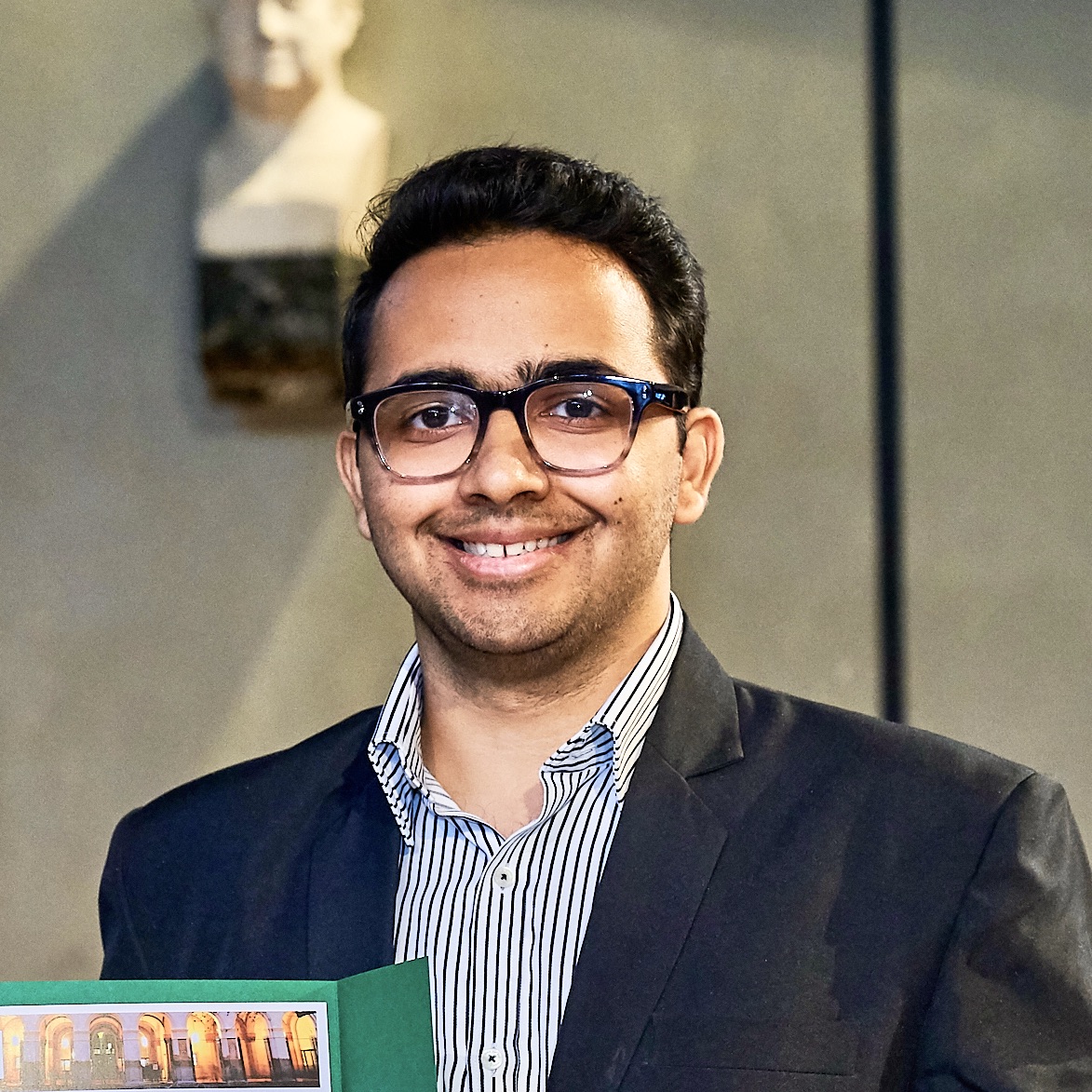}}

\begin{document}
\vspace*{4cm}
\title{An Estimator for Lensing Potential from Galaxy Number Counts}

% \author{Viraj Nistane, Mona Jalilvand, Julien Carron, Ruth Durrer, Martin Kunz}

\author{Viraj Nistane \footnote{\href{https://arxiv.org/abs/2201.04129}{arXiv:2201.04129}, in collaboration with Mona Jalilvand, Julien Carron, Ruth Durrer, Martin Kunz}}

\address{Universit\'e de Gen\`eve, D\'epartement de Physique Th\'eorique and Centre for Astroparticle Physics,\\ 24 quai Ernest-Ansermet, CH-1211 Gen\`eve 4, Switzerland}

\maketitle\abstracts{
We derive an estimator for the lensing potential from galaxy number counts which contains a linear and a quadratic term. We show that this estimator has a much larger signal-to-noise ratio than the corresponding estimator from intensity mapping. This is due to the additional lensing term in the number count angular power spectrum which is present already at linear order. We estimate the signal-to-noise ratio for future photometric surveys. Particularly at high redshifts, $z\gtrsim 1.5$, the signal to noise ratio can become of order 30. Therefore, the number counts in photometric surveys would be an excellent means to measure tomographic lensing spectra.}

\section{Introduction}

Light coming to us from far away sources is deflected by the intervening gravitational field due to cosmic structure which, in the regime of weak lensing and to first order in the cosmological perturbations, can be described by the lensing potential $\phi$.

This work focuses on the measurement of the lensing potential at different redshifts using a new estimator similar to a \textit{quadratic estimator}~\cite{Hu:2001kj,Foreman:2018gnv}.

\section{Galaxy number counts and lensing}

Neglecting large scale relativistic effects which are relevant only at very large scales, the number counts at first order in perturbation theory are given by~\cite{Bonvin:2011bg,Challinor:2011bk}
\be
\label{galaxy1stOrder}
\Delta_{g}(z,\mathbf{n})= b_g(z) \delta -\HH^{-1}\bn\bnabla(\bn\cdot\bV)- \left(2-5s(z)\right) \kappa(z, \mathbf{n}) =\tilde\Delta_{g}(z,\mathbf{n})- \left(2-5s(z)\right) \kappa(z, \mathbf{n})\,.
\ee
The first two terms are the density fluctuation and the redshift space distortion (RSD) which we collect as $\tilde \Delta_{g}$ or $\Delta^{\rm std}_{g}$ as they are also called the `standard terms'. The third term is proportional to the convergence, $\kappa(z, \mathbf{n}) =-\De_2\phi(z, \mathbf{n})/2$, where $\De_2$ denotes the 2D Laplacian on the sphere. The term $2$ in the pre-factor $(2-5s)$ of convergence in Eq.~\eqref{galaxy1stOrder} takes into account the convergence of light rays due to lensing which lowers the number of galaxies per apparent surface area while the term $5s(z)$ accounts for the increase due to the enhancement of the flux in a flux limited sample. Here $s(z)$ is the logarithmic derivative of the number density at the flux limit, $F_*$, of the survey, which corresponds to the luminosity $L_*(z)=4\pi D_L(z)^2F_*$ where $D_L(z)$ denotes the luminosity distance,
\be\label{e:szF}
5s(z,F_*) = 2\left.\frac{\dd\log \bar n(z,L)}{\dd\log L}\right|_{L=L_*(z)} \,.
\ee

While the first order expression is sufficient to compute the variance of the estimator, we want to consider number counts up to second order in perturbation theory for the signal.
At second order (in $\bell$-space and in the flat sky approximation) we obtain~\cite{Nielsen:2016ldx}, 
\bea
\Delta_{g}(\bell,z) &=& \tilde\Delta_{g}(\bell,z)   - \ell^2 \left(1-\frac{5}{2}s(z)\right)\phi(\bell,z) 
\nonumber \\
&& -\int \frac{d^2\ell_1}{2\pi} \tilde\Delta_{g}(\bell_1,z)\phi(\bell-\bell_1,z)\left[ \left(1-\frac{5}{2}s\right)(\bell-\bell_1)^2+\bell_1 \cdot (\bell-\bell_1)\right] \,.
\eea
where we denote
$$
g_\De(\ell,z) = -\ell^2\left(1-\frac{5}{2}s(z)\right) \,\, ; \quad
K_\De(\bell_1,\bell_2,z) = -\left(1-\frac{5}{2}s(z)\right)(\bell_2-\bell_1)^2 -\bell_1\cdot(\bell_2-\bell_1) \,.
$$
In $K_\De$, the second term is the kernel of CMB lensing~\cite{Planck:2018lbu} and intensity mapping lensing~\cite{Foreman:2018gnv}, but the first term is new and only present for number counts. Also new is of course the entire first order term. For the ensemble average at fixed lensing potential this yields
\bea
\lan\Delta_{g}(\bell,z)\ran_\phi = g_\De(\ell,z)\phi(\bell,z)\,\, ; \,\,
\lan\Delta_{g}(\bell,z)\Delta_{g}(\bell',z)\ran_\phi = \delta(\bell+\bell') \tilde C_\ell(z)  - \frac{1}{2\pi}\phi(\bell+\bell') f_\De(\bell,\bell')
\eea
where $f_X(\bell,\bell',z)=K_X(-\bell,\bell',z)\tilde C_{\ell}(z)+K_X(-\bell',\bell,z)\tilde C_{\ell'}(z)$

\section{The {\sc{lin+quad}} estimator}

The expectation value $\lan\cdots\ran_\phi$ is an ensemble average only over any stochastic observable (here, $\Delta_{g}$), at fixed lensing potential $\phi$. This makes sense only if $\phi$ is (nearly) uncorrelated with $\Delta_{g}$.
For sufficiently high redshifts this is usually a good approximation as the lensing kernel peaks roughly in the middle between $0$ and $r(z)$.
We can now derive an  estimator for $\phi(\bL)$ which combines the linear and the quadratic terms in $X$ to which $\phi$ contributes. It is given by
\bea
\hat\phi_\De(\bL,z) &=& A_\De(L,z)N_\De(L,z)\int  \frac{d^2\ell}{2\pi}X(\bell,z)X(\bL-\bell,z) F_\De(\bell,\bL-\bell,z) \nonumber \\
&& + (1 - A_\De(L,z))\frac{X(\bL,z)}{g_\De(L,z)} \label{e:estX}
\eea
where $$F_\De(\bell_1,\bell_2,z) = \frac{f_\De(\bell_1,\bell_2,z)}{2C_{\ell_1}(z)C_{\ell_2}(z)} \quad ,\quad N_\De(L,z) = \left[\int \frac{d^2\ell}{(2\pi)^2}  f_\De(\bell,\bL-\bell,z)F_\De(\bell,\bL-\bell,z)\right]^{-1},$$ $$\textrm{and} \qquad A_\De(L,z) = C_L(z)/\left(g_\De(L,z)^2N_\De(L,z) +C_L(z)\right).$$

By construction $\lan\hat\phi_\De(\bL,z)\ran_\phi =\phi(\bL,z)$. 
Here, imposing that the quadratic part of the estimator is unbiased and has minimum variance allows us to choose $F_\De$ and $N_\De$. Similar conditions for $\hat\phi_\De$ give us the factor $A_\De$.
We have assumed that the $\phi$ power spectrum, which is quadratic in $\phi$, is smaller than both, $C_L$ and $N_\De$ and can be neglected in these expressions.
Note that while the $\tilde C_\ell$'s appearing in $f_\De$ are the theoretical spectra neglecting lensing, those appearing in $F_\De$ are the measured $C_\ell$'s, including both, lensing and (shot) noise. The total noise from the combined linear and quadratic terms then becomes
\be
N_\De^{\rm (tot)}(L,z)  = \frac{C_L(z)N_\De(L,z)}{C_L(z)+g_\De^2(L,z)N_\De(L,z)}\, = \frac{1}{N_\De^{\rm (lin)}} + \frac{1}{N_\De^{\rm (quad)}}\,, \label{e:NtotX}
\ee
where $N_\De^{\rm (quad)} \equiv N_\De$ and $N_\De^{\rm (lin)} \equiv C_L/[L^4(1-\frac{5}{2}s(z))^2]\,$.
\begin{figure}[!h]
    \centering
    \includegraphics[width=\linewidth]{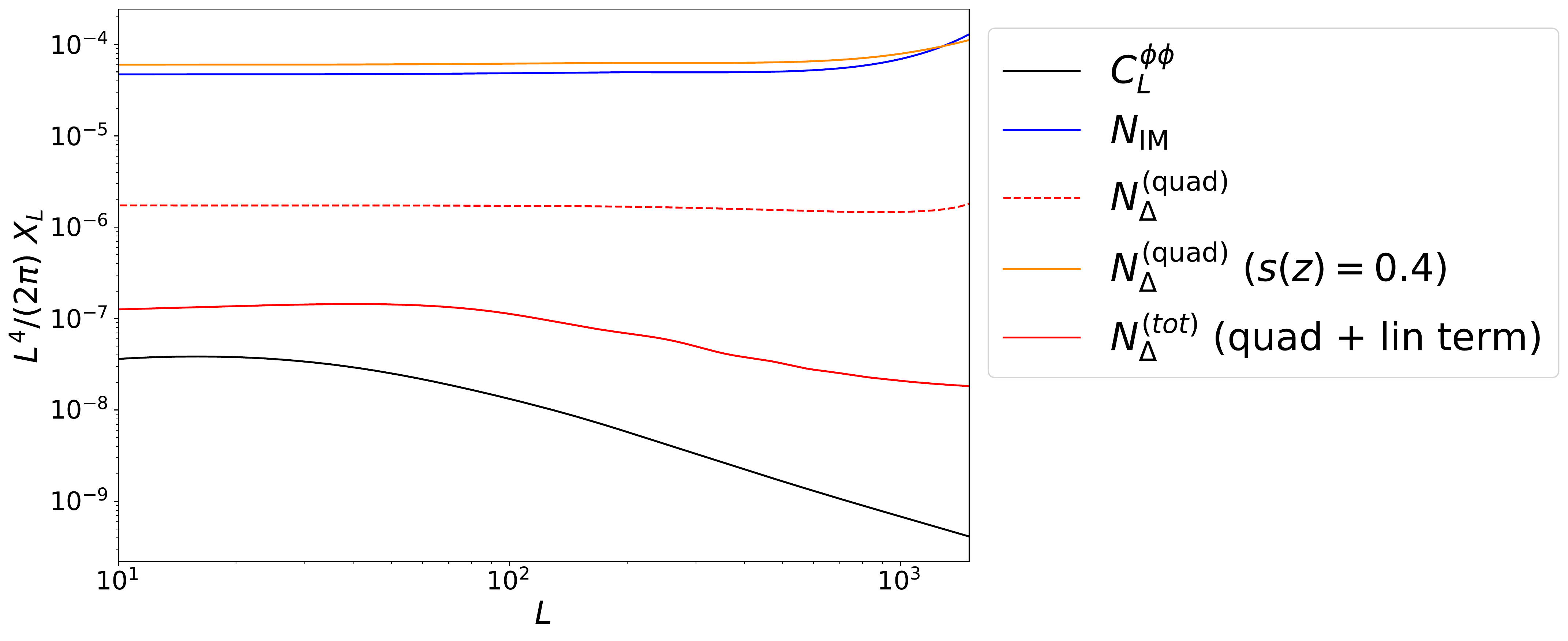}
    \caption{
    The lensing reconstruction noise for halofit power spectra for IM (HIRAX) and galaxy number counts (Euclid-like) for $z=1.9$, $\Delta z = 0.5$. We also indicate the signal $C_\ell^{\phi\phi}$ for comparison. We also show the galaxy number count noise obtained when replacing $s(z)$ by $2/5$. 
    Note that the naive $L^{-4}$ scaling of the noise holds very well for the quadratic noise, but the total noise, $N_\Delta^{\rm tot}$ decays faster for $L>60$. This is due to the significantly smaller linear noise.
    }
    \label{fig:Noise_z_1.91_sw_0.25_linear_galaxy_IM}
\end{figure}

\section{Signal-to-Noise (SNR)}

\begin{figure}[!h]
\centering
\subfloat[]{%
    \label{nz}%
    \includegraphics[width=.49\linewidth]{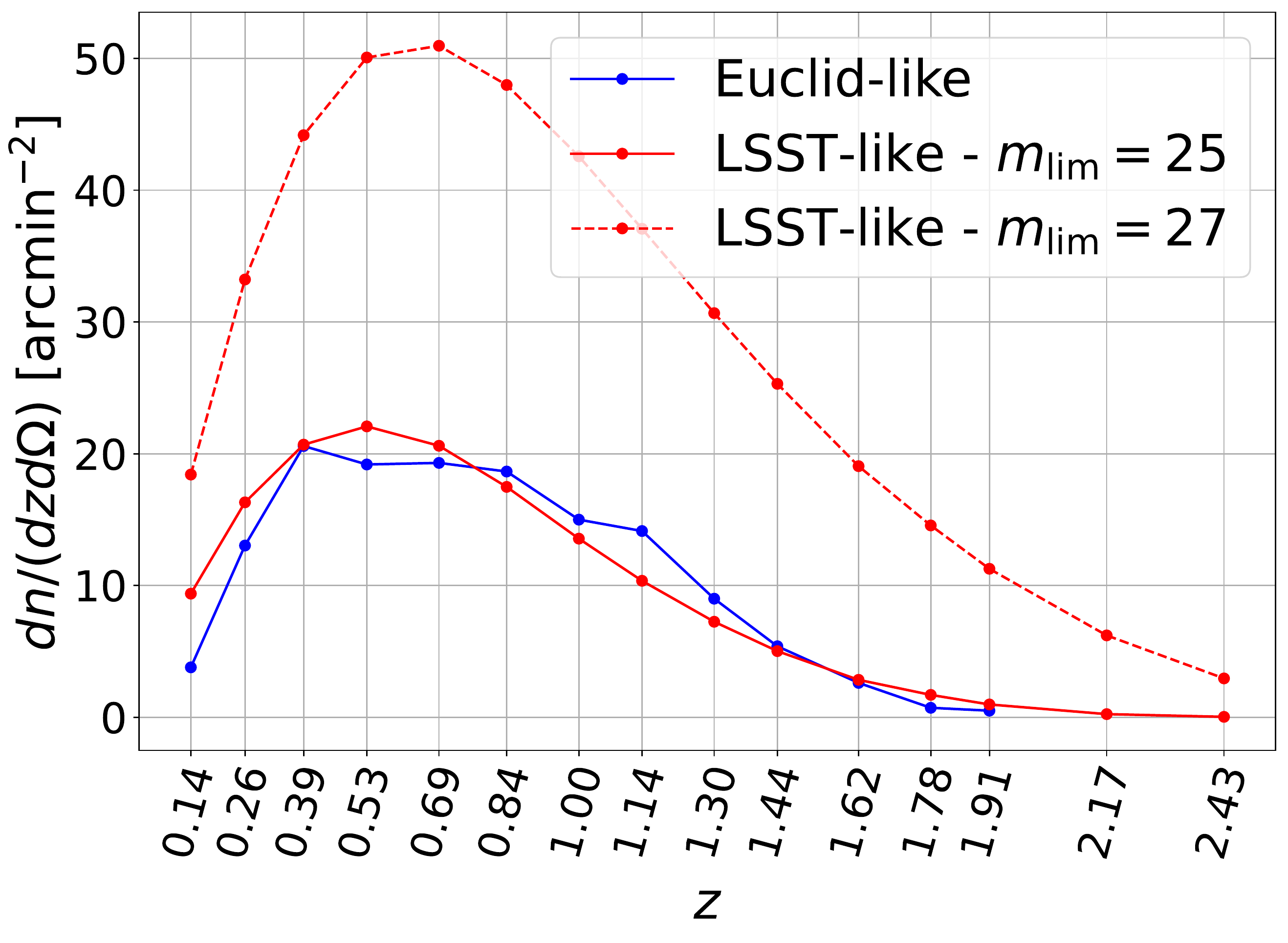}%
}%
\hfill
\subfloat[]{%
    \label{sz}%
    \includegraphics[width=.49\linewidth]{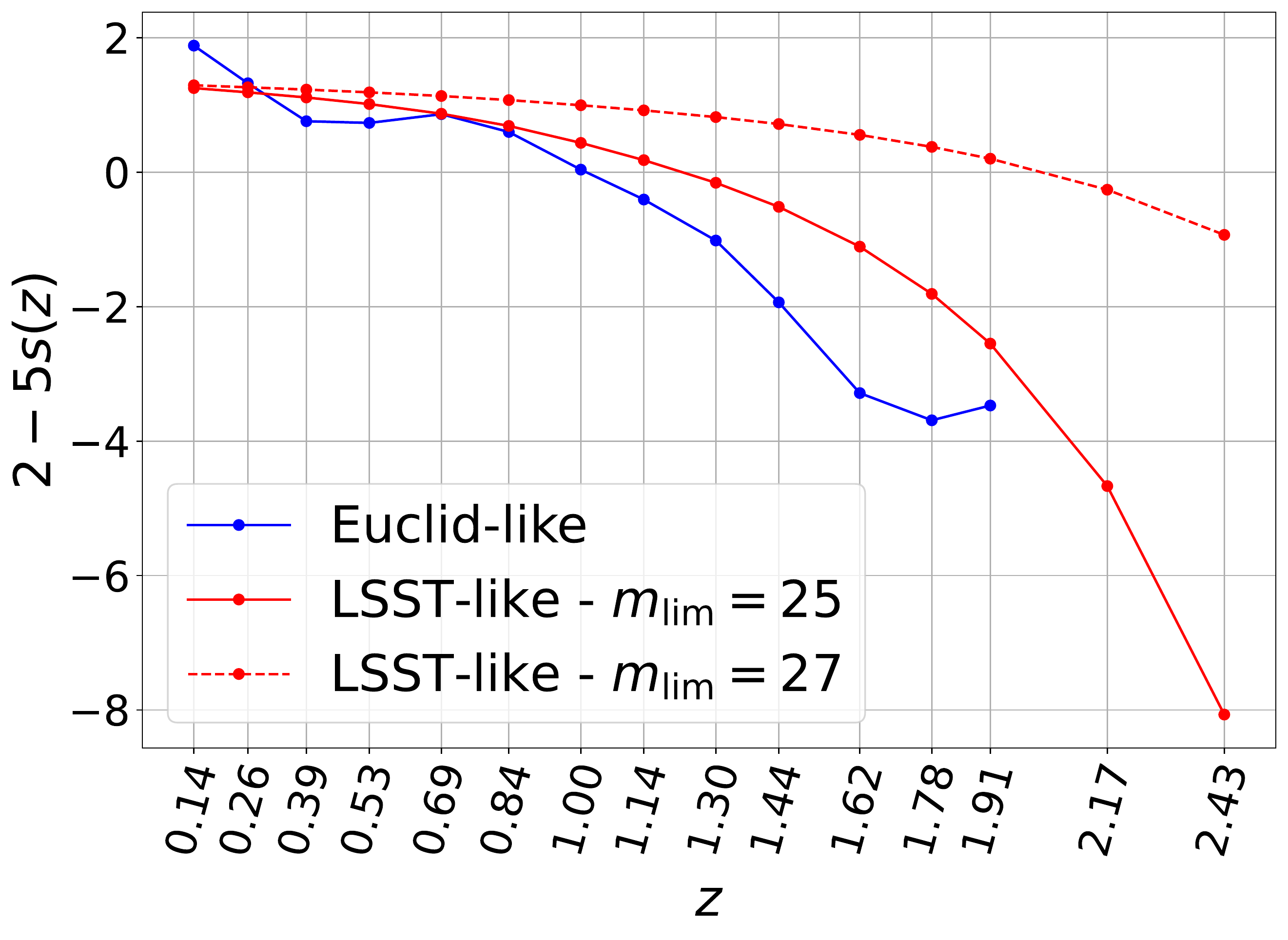}%
}
\caption{\label{nzsz} \protect\subref{nz} {\large{$\frac{dn}{dz d\Omega}$}} [arcmin$^{-2}$], and \protect\subref{sz} forecasts for the magnification bias $s(z)$, for the  surveys~\protect\cite{Alonso:2015uua,Euclid:2021rez} considered in this work.}
\end{figure}

\begin{figure}[!h]
    \centering
    \includegraphics[width=\linewidth]{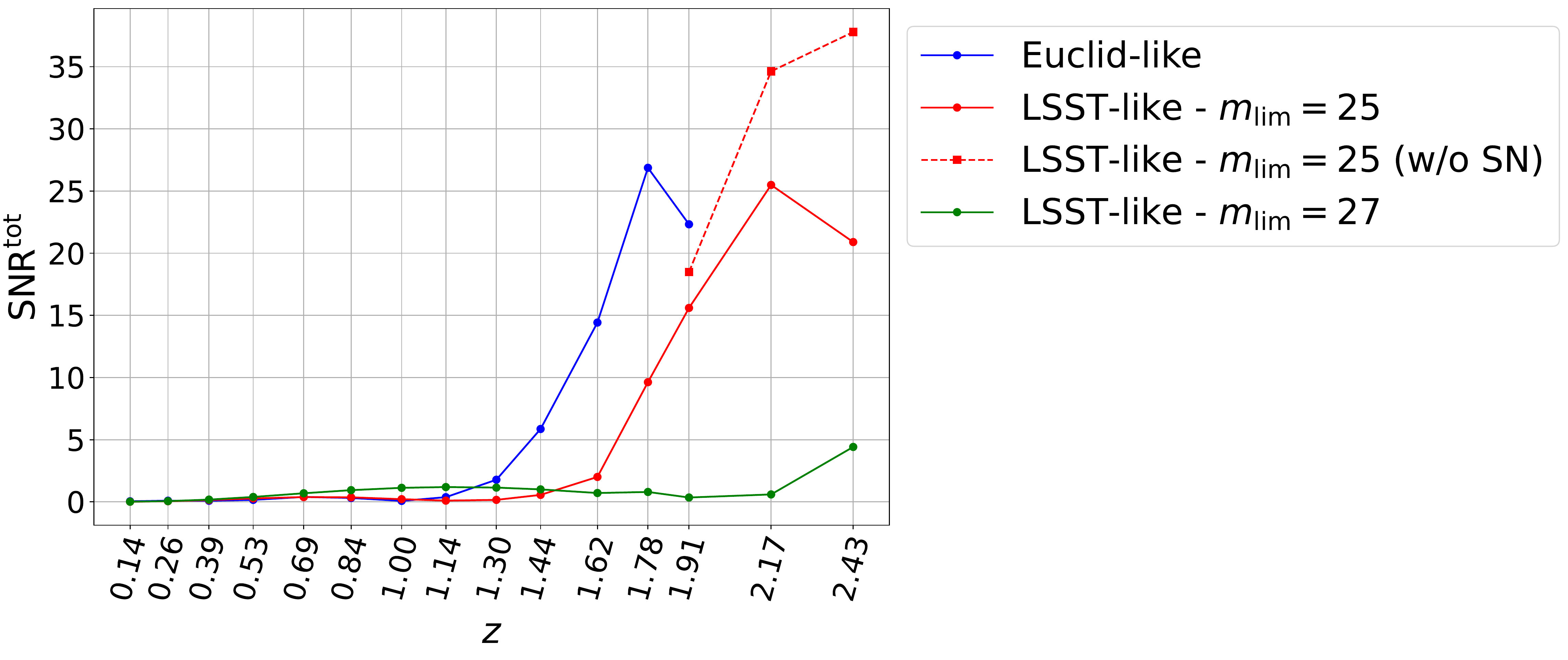}
    \caption{
    Total SNR per redshift bin plotted against the mean redshift of each bin for non-linear perturbation theory results for Euclid-like and LSST-like surveys. In the highest redshift bins of the LSST-like survey, shot noise starts to become important, the red-dashed curve shows the SNR that we would find without shot noise.}
    \label{SNR}
\end{figure}

We consider three exemplary photometric 15'000 square-degree surveys: (1) \textit{Euclid-like} survey~\cite{Laureijs:2011gra} with a limiting depth of 24, (2) \textit{LSST-like-25}~\cite{Abell:2009aa} with limiting magnitude $m_{\rm{lim}}=25$, and (3) \textit{LSST-like-27} with $m_{\rm{lim}}=27$. Forecasts~\cite{Alonso:2015uua,Jelic-Cizmek:2020pkh,Euclid:2021rez} for the number densities $n(z)$ and the magnification bias $s(z)$ are shown in Fig.~\ref{nzsz}. We also use the forecasts for the galaxy bias $b(z)$. The total signal-to-noise values per redshift bin evaluated using Eq.~\eqref{SNReqn} for the estimator are shown in Fig.~\ref{SNR}.
\be\label{SNReqn}
\left(\frac{S}{N}\right)^2(L,z) = 
 \frac{f_{\rm sky}\left(2L+1\right)}{2}\left(\frac{C_L^{\phi\phi}(z)}{C_L^{\phi\phi}(z)+N^{\rm tot}_\De(L,z)}\right)^2;\,\left(\frac{S}{N}\right)_{\rm tot,z}=\sqrt{\sum_{L_{\min}=20}^{L_{\max}=1500}\left(\frac{S}{N}\right)^2(L,z)}
\ee

\section{Conclusion}

We have derived a new linear$+$quadratic estimator for the lensing potential from galaxy number count observations. Contrary to the CMB and intensity mapping, lensing contributes to number counts already at first order in perturbation theory. This leads to us connstructing an estimator for measuring $\phi$ with an additional linear contribution as compared to the quadratic estimator for intensity measurements. As a result, the estimator noise also has a linear contribution which is otherwise absent in CMB/IM. The kernel $K_\De$ of galaxy number count lensing already has an additional term (proportional to $(2-5s)$) which results in the quadratic noise of galaxy number counts estimator being more than an order lower than the quadratic noise (also the total estimator noise) of intensity mapping. In galaxy number counts, the linear noise contribution results in the total lensing reconstruction noise shifting further down by an order. For a high SNR (especially in high redshift bins where the lensing effect is more important), as maximizing $|2-5s(z)|$ is crucial for a high SNR, it may be more optimal in some cases to consider a higher flux limit $F_*$ in order to increase this pre-factor, even though increasing $F_*$  reduces the number density of galaxies and therefore increases the shot noise.

\end{document}